# Release of the Kraken: A Novel Money Multiplier Equation's Debut in 21st Century Banking

*Brian P. Hanley*
*Butterfly Sciences, California*

**Abstract** Historically, the banking multiplier has been in a range of 4 to 100, with 25% to 1% reserve ratios at most layers of the banking system encompassing the majority of its range in recent centuries. Here it is shown that multipliers over 1 000 can occur from a new mechanism in banking. This new multiplier uses a default insurance note to insure an outstanding loan in order to return the value of the insured amount into capital. The economic impact of this invention is calculably greater than the original invention of reserve banking. The consequence of this lending invention is to render the existing money multiplier equations of reserve banking obsolete where it occurs. The equations describing this new multiplier do not converge. Each set of parameters for reserve percentage, nesting depth, etc. creates a unique logarithmic curve rather than approaching a limit. Thus it is necessary to show the behavior of this new equation by numerical methods. Understanding this new multiplier and associated issues is necessary for economic analyses of the Global Financial Crisis.

**Correspondence** Brian Paul Hanley, Butterfly Sciences, 205 El Cajon, 95616 Country, California, USA; e-mail**:** brian.hanley@ieee.org

# 1 Introduction

An invention in banking with calculably greater economic impact than the invention of reserve banking itself has appeared. This invention uses a default insurance note (DIN) on an outstanding loan to move the loan amount from suspense into new capital for banking, sometimes referred to as regulatory capital relief (Sjostrom 2009). The potential for this method appears in US bank examination practice (Division of Banking Supervision and Regulation 2000: 531). This is a type of synthetic capital which has similarity to the authorization by a central bank to classify certain performing loans as valid collateral (European Central Bank 2011). I will show how it radically changes the money multiplier calculations that have been the foundation of reserve banking for hundreds of years. The development of the Kraken money multiplier is shown in Equations 3 through 5, culminating with Equation 5. Since Equation 5 is not reducible, Tables 1–4 show numerical results for multipliers with different parameters and depth of nesting. Figure 3 plots curves on a semi-log graph.

## 1.1 Money-lending

Money-lending is the oldest financing transaction involving money. Money-lending is derived from physical symbols of money in earliest times, where money was made of metal coins, rare shells, and other materials. In this transaction, a party, let us call her Jane, loans money to another party, whom we will call Jack. When Jane loans money to Jack she no longer has the physical money she loaned out, Jack has it. The hope is that Jack will pay Jane her principal back, together with interest. Sometimes Jack may have trouble paying Jane back. To compensate for the risk on her outstanding portfolio of loans, Jane needs to charge high interest. This requirement for high interest raises the risk that loans will not be paid back. High interest also puts limits on the viability of enterprises within an economy dependent on money-lending. In this system, there is no new money creation by lending.

## 1.2 Reserve Banking

Reserve banking is perhaps the most remarkable invention in history. By making loans, banks create newly invented money that is itself deposited into the banking system, and from such new deposits new loans are in turn made. Thus, reserve banking placed into private hands the ability to create money by placing a present value on estimates of future ability to pay. Those private estimates of ability to pay made by bankers have generally proven quite good, with the exception of bubbles that occur irregularly due to "the madness of crowds" (Mackay 2001), referred to in recent years as "irrational exuberance".

Banking evolved from enterprises with the ability to store physical currency safely that enabled money transfers without actually carrying the currency symbol from place to place. These became systems that loaned part of the recorded value of vault storage, over time loaning greater and greater fractions of such deposits held in trust, giving some payment to depositors for use of their money. From these roots was developed our fairly well regulated reserve banking system.

In the modern world, a bank operates based on core capital categorized as tier 1 and tier 2 (Basel Committee on Banking Supervision 2006). In this tier 1 and 2 capital is the money invested by stockholders or other investors along with other instruments. Net earnings from the difference between the interest paid to depositors (or borrowed from another institution) and the interest paid by borrowers is accounted as primary profits for investors in the bank. Tier 1 and 2 capital usually is greater than or equal to capital reserve requirements and is supposed to be secondary in position to the needs of demand depositors. A bank generally loans more money than it has in deposits by maintaining sufficient tier 1 and 2 capital as reserves and thus optimizes profits.

A side effect of modern banking is that it makes it possible for banks to charge what we now consider reasonable interest rates while being more profitable than money-lenders are and carrying lower overall risk. This is due to the difference between the borrower's loan repayment income streams relative to core investor capital and the income of the simple system of money-lending. In banking, overall risk is lowered dramatically vis-à-vis money-lending both because lower interest rates are less likely to precipitate default and because a much larger pool of borrowers exists relative to invested capital.

### 1.3 Money Multiplier in Reserve Banking

The standard formula for the banking money multiplier, ***m*** is:

$$m = \frac{1}{R} \tag{1}$$

where ***R*** = capital reserve fraction

The primitive equation is below. At the limit, it renders to the simple one above.

$$m = \sum_{i=1}^{n} (1 - R)^i \tag{2}$$

where ***R*** = capital reserve fraction
***i*** = iteration number on loans/deposits
***n*** = iteration limit

Equation 2 has an asymptote at Equation 1.

In Figure 1 is a curve relating the number of iterations of Equation 2 with the multiplier achieved for that iteration. In a hypothetical system, if 30 days are required to approve each loan after acquiring new capital, then in one year 12 iterations are possible in that system. However, the number of iterations of loans for a real world banking system is variable with regard to time and this 30 day hypothetical system is simply a model system.

In the real world, a bank's lending is dependent on both capital availability and borrower creditworthiness which are both dependent on factors beyond the scope of this discussion. The iteration period could be short if a backlog of approved borrowers is present, making the multiplier high in a short time period. Conversely, if a bank lacks creditworthy borrowers then there is no multiplier at all. So, in the real world, the multiplier can be extremely variable versus time; there is no rule that adding X amount of capital to banks will result in Y amount of new money created in any fixed time period.

In addition, in the real world, money is also taken out as circulating cash, loan losses, etc. so the true multiplier is always less than the theoretical values given by Equation 1 or 2. Similar considerations apply to the equations developed below to describe the multiplier that will be described.

*Figure 1*: Iteration (x axis) versus multiplier (y axis) for a 5% reserve banking system

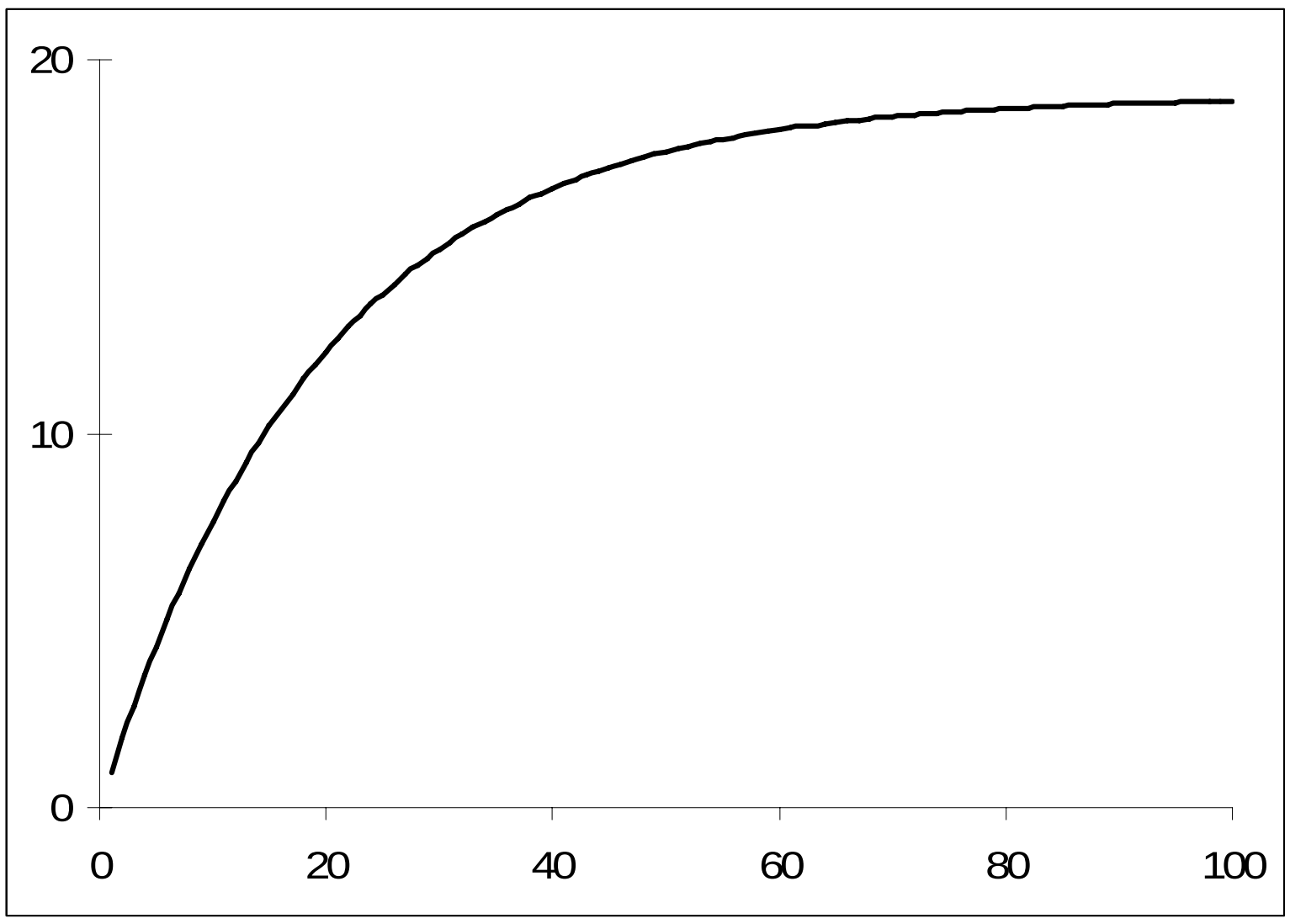

## 2 Kraken Equation Development

Much recent attention in banking centers on what can be allowed on the books as capital for reserve purposes and capital reserve levels. For instance, capital raised by use of trust-preferred securities has been a bone of contention for regulators as is reflected in legislation (Frank, B. and Dodd, C. J. 2010). Some banks have gotten unusually low reserve requirements approved in the recent past leading to regulators demanding increases in reserve ratios (Hawken, K. and Bake, M. 2009). In those discussions, the fundamental principle of the money multiplier in banking is not considered at risk. However, there is a class of capital for which the standard money multiplier formula fails. To calculate the correct multiplier for this class of capital requires a new equation be constructed from the bottom up. Figure 2 shows a schematic diagram of this potential new capital creation.

*Figure 2*: New capital creation schematic for Equations 3–5.

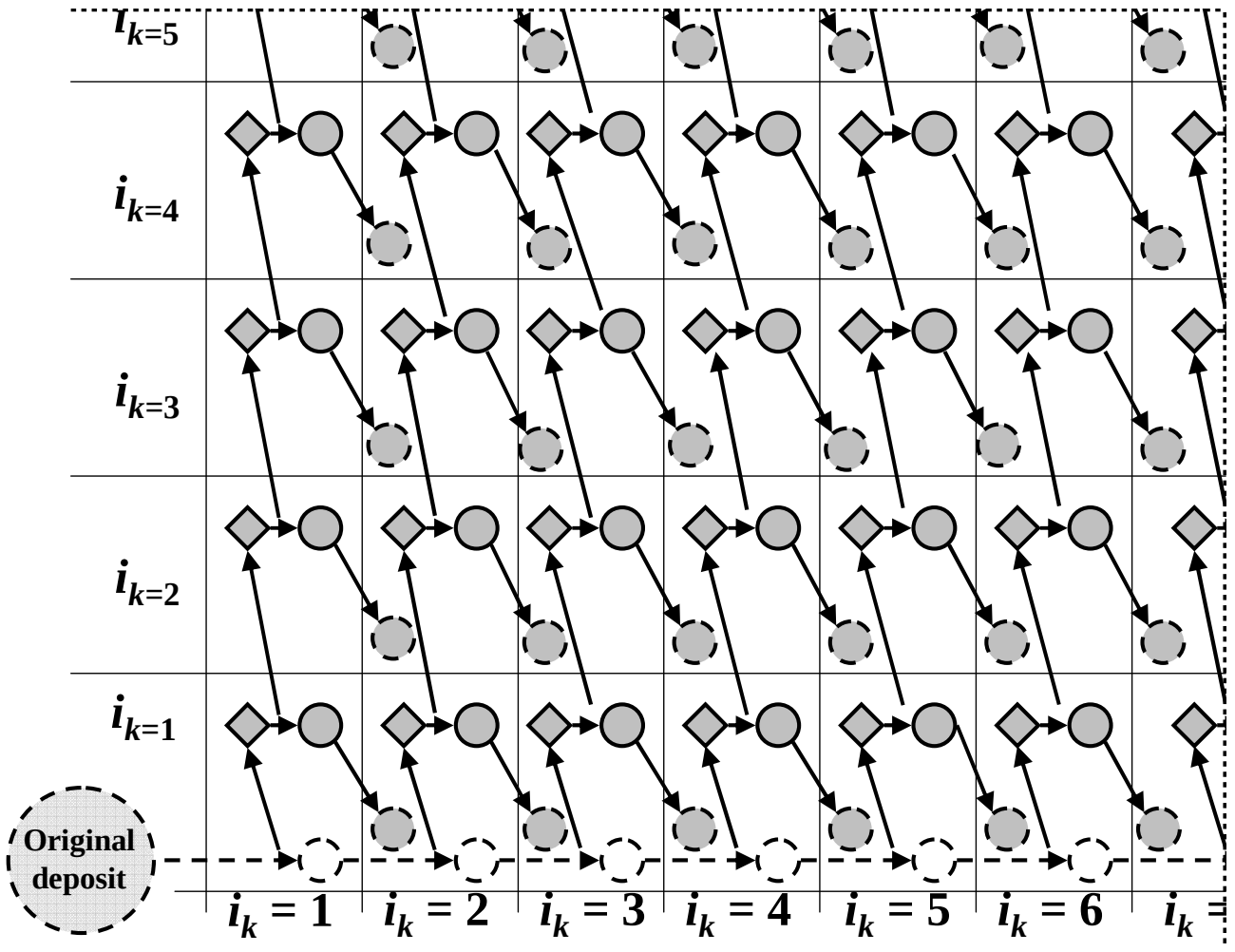

Dashed arrows and dashed circles represent money creation in the normal reserve banking system. Horizontal dashed arrows are the first ***(1–R)***$^{i}$ sub-term of Equations 3–5.

Dashed circles are new loans which become new capital deposits in the standard manner.

Solid upward diagonal lines are the ***(1–I)*** sub-term for acquisition of a DIN based on each loan. These are the ***(O–I)*** term in Equations 4 and 5.

Solid gray-filled diamonds are the new capital created by each DIN acquisition.

Solid horizontal arrows from diamond to circle are the second ***(1–R)***$^{i}$ sub-term.

Solid gray-filled circles are new loans made which become new capital deposits.

Downward diagonal arrows are the first new ***(1–R)***$^{i}$ sub-terms.

Dashed gray-filled circles are new loans made from new loan deposits in the conventional way, each of which anchors a new schematic identical with this one as an "original deposit". However, for diagrams past the first layer shown here, the dashed circle amounts of the first line are no longer part of the conventional banking multiplier system.

## 2.1 Equation 3—Nested *k* Depth Iteration with Insurance Cost

This novel invention of banking is based on the concept of acquiring a default insurance note (DIN) for an outstanding loan provided by a bank. The default insurance note guarantees to the bank that the loan will be paid off should the borrower go into default. In the recent scenarios of the GFC, these were provided by credit default swaps (CDS's) provided primarily by AIG on real estate loans. The bank pays a premium (nominally 1% to 5%) and receives in return a DIN for a loan that has been issued. On the basis of the DIN received, the bank puts the value of the note (or some major fraction thereof) into its capital account and can then write a new loan. This is described by Equation 3.

$$m=\sum_{i_1=1}^{n}((1-R)^{i_1}+((1-R)^{i_1})(1-I)\cdot\sum_{i_2=1}^{n}((1-R)^{i_2}+((1-R)^{i_2})(1-I)$$
$$\cdot\sum_{i_{k\ldots}=1}^{n}((1-R)^{i_{k\ldots}}+((1-R)^{i_{k\ldots}})(1-I)\ldots)))))) \tag{3}$$

Where: ***R*** = deposit reserve fraction,

$\boldsymbol{i_1}$, $\boldsymbol{i_2}$, $\boldsymbol{i_{k\ldots}}$ = iteration number on loans/deposits, ***k*** being the series end term

***n*** **=** iteration limit

***I*** = insurance price as fractional cost

A default insurance note would be paid over time. The nominal cost is based on a reported 5 year term for the CDS with quarterly payments of 0.5%. Very low premiums were reported as AIG's CDS business increased. In essence this created the opportunity to rent capital that could in principle be rolled over more or less indefinitely. (However a higher cost was used in calculations for conservatism.)

In this new scenario described by Equation 3, not only does each loan become new capital when the loan is deposited into the banking system by the borrower, but in addition, each DIN allows booking of brand new synthetic capital for the bank that holds the loan. Then, the synthetic capital becomes the basis for a new loan and, that new loan becomes a new deposit of ordinary capital somewhere. In turn, each of those new deposits of capital becomes the potential basis for a new loan until the fractional capital available peters out. As shown in Figure 2, in this new scenario money multiplies more than geometrically. It is completely different than the simple assumptions underlying Equation 2. Mathematically this is very interesting. It is a nested summation that can show hyper-exponential behavior depending on parameters (for Equations 3–5).

There are several new things in Equation 3 compared to Equation 2. There is the second sub-term adding a new loan. Within this sub-term are ***I***, the price of the default insurance note, and its companion term for allocating a new loan from new capital. And there are the series iteration summation terms that are multiplied in a nested manner. The equation is remarkable; it is a nested multiplicative series of summations. When historical values are plugged in of ***R*** = 5% and ***I*** = 5% we get very interesting numbers, as will be seen.

Restated, for each loan made, a default insurance note is acquired. This DIN is used to declare that the loan amount minus the cost of the DIN is new capital. Each default insurance note is then used to originate a new loan, which loan gives rise to more synthetic capital, etc.

## 2.2 Equation 4 – Addition of Origination Fee

Examining Equation 3, it becomes apparent that an assumption with possible impact on the multiplier equation's behavior may not be entirely correct. In Equation 3, the cost of ***I*** is subtracted from 1, which assumes the starting point is the value of the loan being insured with a DIN. However, banks charge origination fees (points) to borrowers. If all or part of the cost of ***I*** is paid for by such fees, then the subterm **( 1 - *I* )** becomes **( *O* - *I* )** where ***O*** is loan plus origination fees and greater than or equal to 1. If  is possible to charge more in fees to originate the loan than the cost of ***I***, then the same subterm can even evaluate to greater than 1. This generates more interesting behavior as shown in Equation 4.

$$m = \sum_{i_1=1}^{n}((1-R)^{i_1} + ((1-R)^{i_1}(O-I)\cdot\sum_{i_2=1}^{n}((1-R)^{i_2} + ((1-R)^{i_2}(O-I) \cdot\sum_{i_{k\ldots}=1}^{n}((1-R)^{i_{k\ldots}} + ((1-R)^{i_{k\ldots}}(O-I)\ldots)))))) \quad (4)$$

Where: ***R*** = deposit reserve fraction,

$i_1$, $i_2$, $i_{k\ldots}$ = iteration number on loans/deposits, ***k*** being the series end term

***n*** = iteration limit, ***I*** = insurance price as fraction

***O*** = 1 + origination fee fraction of loan (generally charged as "points")

Origination fees of 5 points are about as high as such fees go and 5% origination fees are reasonable to view as limits under normal circumstances.

## 2.3 Equation 5 – Addition of Tranche Fraction

In examining Equation 4, a final parameter becomes visible, which is ***T***, the tranche fraction for a portfolio for which a DIN is obtained. This results in Equation 5 shown below. In Tables 1–4 is graphically shown how multipliers grow with nesting of Equation 5 up to 10 levels deep for a realistic insurable tranche fraction of 30%.

$$m=\sum_{i_1=1}^{n}((1-R)^{i_1}+(((1-R)^{i_1}(O-I)\cdot T)\cdot\sum_{i_2=1}^{n}((1-R)^{i_2}+(((1-R)^{i_2}(O-I)\cdot T)$$
$$\cdot\sum_{i_{k\ldots}=1}^{n}((1-R)^{i_{k\ldots}}+(((1-R)^{i_{k\ldots}}(O-I)\cdot T)\ldots)))))) \quad (5)$$

Where: ***R*** = deposit reserve fraction,

$i_1, i_2, i_{k\ldots}$ = iteration number on loans/deposits, ***k*** being the series end term

***n*** = iteration limit, ***I*** = insurance price as fraction

***O*** = 1 + origination fee fraction of loan (generally charged as “points”)

***T*** = tranche fraction insured.

It is instructive to see what happens using Equation 5 when ***T*** = 0.30, ***O*** = 1, ***I*** = 5%, ***n*** = 100, and ***k*** increments from 1 to 10 (e.g. a maximum of 10 levels of nested iterations) with different values of ***R***. These are shown in Tables 1 and 2 below.

It is further instructive to see what happens using Equation 5 for the same cases as tables 1 and 2 when ***O*** = 1.05 to account for fee level of 5 points on origination of the loans. This is shown in Tables 3 and 4 and in Figure 3.

*Tables 1 and 2*: A sample of multipliers for values of ***k*** from 1 to 10 are presented

| Table 1 | | Table 2 | |
|---|---|---|---|
| k | ***m, where T=0.30 O=1, R=0.05*** | k | ***m, where T=0.30 =1, R= 0.025*** |
| 1 | 24 | 1 | 46 |
| 2 | 150 | 2 | 508 |
| 3 | 824 | 3 | 5,232 |
| 4 | 4,453 | 4 | 53,565 |
| 5 | 23,992 | 5 | 548,064 |
| 6 | 129,164 | 6 | 5,607,368 |
| 7 | 695,302 | 7 | 57,369,941 |
| 8 | 3,742,788 | 8 | 586,961,390 |
| 9 | 20,147,225 | 9 | 6,005,299,050 |
| 10 | 108,451,327 | 10 | 61,441,207,420 |

Values of ***m*** for increasing ***k***, when ***n*** =100, ***O*** =1.0, ***I*** =5%, ***T*** =30% for ***R*** =5% and ***R*** =2.5%. Numerical methods used because equations do not converge and are not reducible.

*Tables 3 and 4*: A sample of multipliers for values of ***k*** from 1 to 10 are presented

| Table 1 | | Table 2 | |
|---|---|---|---|
| k | ***m, where T=0.30 O=1.05, R= 0.05*** | k | ***m, where T=0.30 =1.05, R=0.025*** |
| 1 | 24 | 1 | 46 |
| 2 | 158 | 2 | 538 |
| 3 | 914 | 3 | 5,835 |
| 4 | 5,199 | 4 | 62,880 |
| 5 | 29,479 | 5 | 677,240 |
| 6 | 167,054 | 6 | 7,293,674 |
| 7 | 946,589 | 7 | 78,550,336 |
| 8 | 5,363,637 | 8 | 845,959,488 |
| 9 | 30,391,743 | 9 | 9,110,685,705 |
| 10 | 172,207,323 | 10 | 98,118,875,480 |

Values of ***m*** for increasing ***k***, when ***n*** =100, ***O*** =1.05, ***I*** =5%, ***T*** =30% for ***R*** =5% and ***R*** =2.5%. Numerical methods are used because equations do not converge and are not reducible.

*Figure 3*: Graph of Equation 5 for reserves of 5% (red) and 2.5% (blue)

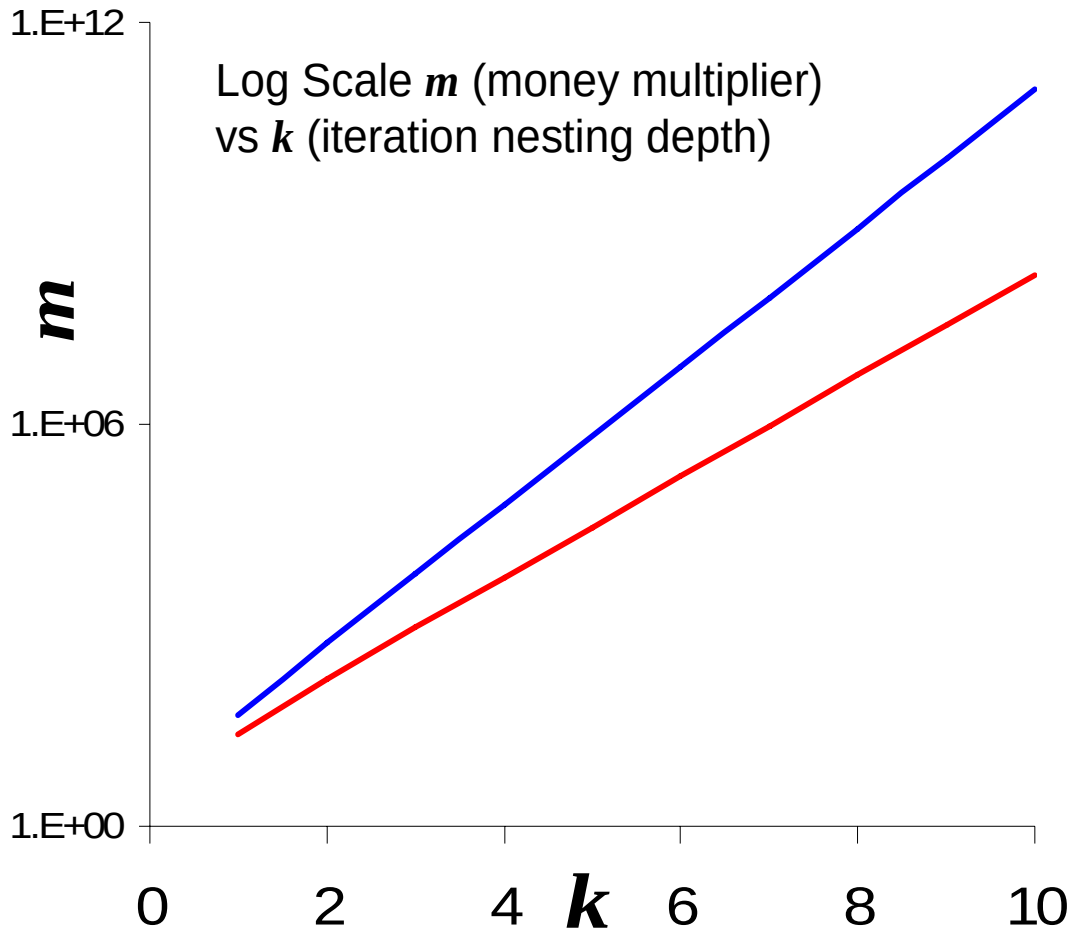

Shows values of ***m***, for ***k*** from 1 to 10, when ***n*** =100, ***O*** =1.05, ***I*** = 5%, ***T*** = 30%.
(Red) ***R*** = 5% (Blue) ***R*** = 2.5%.
Each set of parameters will create a unique logarithmic curve rather than approach a limit. It is obvious from inspection of this figure that while the Kraken mechanism can be used to create money through credit without any internal limit, this mechanism must bump up against limits to money creation present in the outside world.

The point of showing the numerical results in Tables 1–4 and Figure 3 is to illustrate that there is no reasonable limit to money creation using this mechanism inherent to its mathematics. This is a major departure from the money multiplier mathematics that has heretofore been understood. Each set of parameters results in a unique logarithmic curve. Therefore, Equation 5 can, in theory, create billions of dollars for every original dollar. The only significant limitations are: A.) a bank's ability to acquire default insurance note contracts and if necessary maintain them by rollover; B.) to some degree limits due to regulatory tier 1 and 2 capital type restrictions where BIS standards are observed, which is discussed below; and C.) external limits on credit creation, also discussed below.

## 3 Discussion

Credit default swaps were introduced in the mid-1990's and became more widely used for pure speculative and synthetic capital activities later. AIG is an example of a monoline issuer as opposed to banks which also issue such instruments (Weistroffer 2009). There are a number of forms of CDS, and variants tailored for specific circumstances. This analysis assumes the most common physical settlement method which delivers the underlying security in return for face value of the CDS. Variants would need to modify these equations appropriately.

Most historical uses of credit default swaps have nothing to do with synthetic capital creation, but with risk management (Sjostrom 2009; Weistroffer 2009). This discussion should not be construed as inherently critical of CDS instruments but of their use outside a sound system of credit evaluation and transparency.

CDS instruments present special issues. The relationships between an insurer and the insured can be rather complicated. Company B may buy a CDS from issuer A, but company B may in turn sell a CDS to company C for a premium over the CDS from issuer A. Company C may in turn issue a CDS to company D in the same manner. The chain tends to extend as market conditions get progressively worse, since premiums rise under those conditions and incentives to cash in on a premium increase. As CDS instruments are unregulated, it is not generally possible to know who the terminal counterparty is. A party failing anywhere within the chain may have knock-on effects for other parties.

For instance, in the above chain, if issuer A fails, then company B is saddled with the full liability. If company C fails, then company D is on the hook to pay, and whether company B is on the hook depends on how transparent company C's affairs are and the time scale of collections.

This suggests that the CDS market could be improved by stating, within the contract, whether the seller is the initial issuing party. When a CDS is sold based on an underlying CDS, the underlying CDS chain of counterparties could be listed on the instrument tracing back to the originator. Adding the obligation that CDS originators inform the chain if the originator buys a CDS to cover its own risk would also be needed to implement such a policy.

The Kraken multiplier mechanism is not inherently problematic since in theory default insurance notes should only be available for properly rated loan-backed securities. Consequently, in a regulated environment where security risk rating in aggregate was adequately done, and the economy needed the capacity to create large amounts of money in order to create new utility value, this new mechanism could have a perfectly valid place. There is little question of the economic utility to private money creation in private hands through banking. This mechanism, in theory, should simply improve matters.

However, in practice, the short term rewards of the Kraken mechanism and other aspects of derivative use made bypassing proper rating of securities irresistible to bankers (The Financial Crisis Inquiry Commission 2011). Thus, a bubble appeared, running up prices in a speculative price-kiting-economy. The results were dramatic, and pondering this mechanism suggests that deliberate exploitation could prove catastrophic. This suggests that transparency and improvement of regulatory oversight of the use of CDS's would be appropriate. Details of these aspects are discussed below.

## 3.1 Limits on Loan Insurance Providers Ability to Cover Losses

An insurer must have a model of risk involved in selling policies in order to be viable long-term. In the specific case of AIG's CDS business, the empirical historical model used was inadequate, as it did not accurately rate the risk of default within the real estate loan market.

It can be argued that this failure to accurately assess risk was inevitable, since CDS instruments were new and without an historical example of large-scale default within this sector, an empirical model could not be sufficient. Opposing this argument is the history of real estate price booms, busts, and plateaus which should indicate a need for combining empiricism with theoretical models. However, models that show limits to growth present modelers with what is, arguably, a need to estimate the top of the market long before it happens. In the ongoing activity of a market with strong incentives to continue, ending profit-making prematurely is not a popular activity.

Thus, there is a problem with properly estimating risk, and it is noteworthy that market pressure to maintain what may be invalid assumptions rises with the real risk level. In a loan market in which a bubble is developing, activity accelerates, and as it accelerates, greater apparent profits are seen in the short term. Profitability in the near term can be very difficult for a modern corporation to argue with. Thus, a Ponzi economy occurring within some sector has strong disincentives to accurately assess risk.

Further complicating the assessment of risk are incentives to unrealistic ratings issued by credit rating agencies (Levin, C., *et al.* 2011). Since an insurer cannot trust that CRA ratings are necessarily accurate, then to rate risk properly, perhaps establishing that rating agencies for CDS use must be paid by the buyers of CDS instruments rather than providers of the underlying financial instrument may prove effective.

Significantly complicating the assessment of risk is material that has come to light as a result of SEC prosecution. For instance, Citigroup is accused of lying to investors who bought securities, claiming they were selected by an independent agent. Those securities went into default, and Citigroup had bought CDS's against the underlying securities (Rakoff 2011; Wyatt 2011). It is easy to see that it is more profitable for a bank to make a poor quality loan, buy a CDS, sell off the low quality loan under false representation, then collect on the CDS, than it would be to simply make the loan and collect interest on it after fees. This is a strong incentive to economically pathological banking behavior.

This pathological complication of modern finance represents a serious concern for evaluation of risk by insurers. Buying a CDS on a security which is believed by the buyer to be soon insolvent is, at best, a relationship in which good faith does not exist. The buyer knowingly withholds information critical to the seller. In those cases no actuarial model applies. For the initial issuer, such a sale has similarities to an insurer selling a fire insurance policy to an arsonist or life insurance to a person intending suicide. Like arson or suicide, determining intent or knowledge of the buyer can be difficult.

It would seem that for buyers of CDS instruments the ability of an insurer to pay should be part of their trading model. Thus, at least in a system without publicly funded rescue, rational players would limit the number of bad-faith CDS acquisitions in order to ensure that they would be paid off. However, in a system which is lacking in transparency, it is difficult to determine what the liabilities of major CDS issuers are.

Being unable to determine net liabilities of a CDS issuer creates serious problems, even in a system without publicly funded rescue. Where it is not practical to determine the liabilities of a CDS issuer and rational players are aware that the issuer must collapse eventually, a rational player should maximize bad-faith profits in the short term in order to beat other players to the payoff before the insurer is unable to cover its losses.

Within a “too big to fail” system, wherein losses by the insurer will be covered by the public purse, a rational player should have little incentive toward limitation, whether the system is transparent or not. Instead, a rational player should attempt to carry out as many bad-faith transactions as possible in order to maximize profits. The only concern should be over when the clock will run out.

There are some disincentives for failure in the AIG scenario. CEOs were replaced and some withholding of compensation occurred (Dennis 2009; Zuill 2008). But the size of short term financial incentives probably overrules such disincentives.

For the purposes of this paper, this subsection describes how special pathological incentives exist that should lead to breakdown of the default insurance note synthetic capital mechanism. In any case, an insurer operating in a bubble environment is guaranteed to have losses that cannot be covered.

### 3.2 Special Treatment of CDS and Other Derivative Instruments in Bankruptcy

First appearing in the United States in 1982, safe-harbor bankruptcy code provisions put the claims of derivative holders first, granting them the right to terminate and complete transactions immediately upon bankruptcy of a counterparty (Gilbane 2010). Additionally, derivative holders have the right to immediate foreclosure on underlying assets. These safe-harbor protections cover forward, commodity, and security contracts, repurchase and swap agreements. All other claims in bankruptcy are given an automatic stay and must wend their way through the courts. These provisions were clarified and strengthened in 2005 to ensure newer instruments would be covered with the passage of public law 109-8, or BACPA (Grassley 2005). These provisions of law are now in place in most of the world.

The impact of special treatment in bankruptcy on the current thesis is indirect, but it is useful to understand these provisions in order to properly comprehend how such instruments work and the motivations surrounding them. The recent collapse of MF Global illustrates a potential problem where such instant claim resolution may result in removal of assets to which the creditor has questionable claim (Alper and Viswanatha 2011; Alper et al. 2011). That problem must then be resolved through the courts. The relative time scale of these methods of resolution can have side effects that are quite hard to predict.

One proposed side-effect of current law is that it may in some cases encourage counterparties to desire the bankruptcy of asset holders and perhaps take steps to foster such. Such a pathological motivation to insolvency in a financial system using Kraken banking make it quite unlikely to be metastable.

## 3.3 Tier 1 and Tier 2 Capital Limits versus Creation of Demand Deposits

Noting first that the definitions of capital categories are flexible, it will be observed that direct DIN capital would be defined as tier 1 up to the 15% rule per Basel II (Basel Committee on Banking Supervision 2006), after which DIN capital would become tier 2 capital, and hence subject to tier 2 limitations. It does not appear that this rule has changed for Basel III ( Basel Committee on Banking Supervision 2010). Within the banking system as a whole, any loans created from such presumptive tier 1 or 2 capital will represent new demand deposits to the banking system and these loans would be in superior position to tier capital. Any secondary loans granted from the deposits using the standard money multiplier mechanism will also be demand deposits within the system as a whole. In theory, it is possible that demand deposits could be created so as to require retention of such monies to meet reserve requirements.

Examining Equation 5, it can be seen that for each term, the subterm quantity representing the DIN capital for its corresponding loan is lower than the default insurance note itself by the amount of the retained reserve percentage. This can be used to yield an Equation 6 which indicates that the ratio between logged DIN capital and new deposit creation from the mechanism itself.

As long as the assumption is made that each new loan is in turn insured, there will be some limits to short term capital creation because $r_{DIN}$ is greater than one, and so tiered capital limits will eventually be exceeded. Over longer time scales, circulation of created money could result in enlarging the pool of tier 1 and 2 investment capital. However, tiered capital only becomes a limit if the bank does not increase its reserves through usual fiat money mechanisms.

$$r_{DIN} = \frac{1}{(O-I)\cdot T} \tag{6}$$

Where: $\boldsymbol{r_{DIN}}$ = ratio of DIN capital to new deposit creation
$\boldsymbol{I}$ = insurance price
$\boldsymbol{O}$ = 1 + origination fee fraction of loan (generally charged as "points")
$\boldsymbol{T}$ = tranche fraction insured (See Equation 7.)

In a 5% reserve system $\boldsymbol{r_{DIN}}$ will be approximately 1.052 per transaction.

However, in cases in which some fraction of new loans are not insured the ratio drops below 1 as shown in Equation 7.

$$r_{DIN} = \frac{1}{(O-I)+\sum_{i=s}^{n}(1-R)^{i}} \tag{7}$$

Where: $\boldsymbol{r_{DIN}}$ = ratio of DIN capital to new deposit creation
$\boldsymbol{I}$ = insurance price
$\boldsymbol{O}$ = 1 + origination fee fraction of loan (generally charged as "points")
$\boldsymbol{i}$ = iteration number on loans/deposits
$\boldsymbol{s}$ = iteration start, $\boldsymbol{n}$ = iteration limit for each loan not in turn DIN insured

In a simple example case of skipping one loan for DIN insurance, the ratio for that pair in a 5% reserve system would be approximately 0.54. Such a ratio would grow total deposits in the system at almost double the rate of the recording of DIN tier 1 or 2 capital. It could be theoretically possible that deposits could increase to a level such that demand deposit monies would be required to be retained to maintain capital reserve needs. In such a scenario, the limits on money creation would be unclear.

Of course, the banking system is not homogenous, but is composed of institutions that exhibit degrees of source and sink characteristics for any particular financial instrument. Consequently, should a limited number of institutions be generating new money by the DIN mechanism, that set of institutions will be limited in their money creation to the extent that new loan deposits leak from their institutional group. This leads to the conclusion that limits on creation of default insurance note originated money can vary greatly on a case by case basis. But it would appear likely that DIN capital creation would experience its first realistic limit in the crash which would generally follow an overexpansion of such capital based on any particular narrow hard-asset class. This could perhaps be prevented by using such a mechanism only with a combination of hard assets and enterprises creating new utility value.

### 3.4 Bailouts and the Creation of a Bubble

The mechanism of Equation 5 creates money as fast as the transactions can occur. It is a mechanism that generates money on a scale that would rapidly eclipse the previously created monetary wealth of the world if it were possible for it to operate unchecked for long.

"Too big to fail" bailouts, (of Long Term Capital Management, then of AIG and a host of banks) has privatized profit and left the public holding the bag for losses due to risk, thus undermining fundamental efficient market assumptions (Hetzel 1991; Quiggin 2010: 50–51). The mechanism of bubble creation relative to DIN banking is presumed to be that it encourages lenders to ignore credit risk (Dickinson 2008). This relies on failure of rating agencies to inform DIN sellers as well as failure of risk assessment by the sellers and these elements are present in the GFC (The Financial Crisis Inquiry Commission 2011).

When an asset such as real estate turns over for an increased price, little or no new utility value is normally created. Creation of new utility value occurs if the home is built and sold for the first time where there is need for the housing, or to the extent it is improved commensurate with the price increase when turned over. Radical price increases for homes in the years just prior to the bubble were seen in the USA before the real estate bubble popped.

In an ideal model system, new capital would be accompanied by the creation of new value of goods and services commensurate with it, so the amount of money available to purchase does not rise dramatically while supply remains fixed (thus inflating price). Without commensurate increase in utility value of goods, increases in price due to excessive money supply creates a bubble and the market resembles that for tulip bulb mania. So a DIN mechanism that exists only for a narrow sector would be expected to create a bubble by price inflation as the mechanism created more money in that sector. Thus, the mechanism of DINs applied to real estate loans would be expected to create a bubble due to the massive amount of new money that can be created entirely focused on a narrow fixed asset class.

### 3.5 Lower Reserves and Raised Fees in Concert with Use of the Kraken Mechanism

In Table 1 we see that the hoary reserve multiple of Equation 1 has increased by 20% from 20 to 24 at the first iteration ( $\boldsymbol{k}$ = 1 ) using Equation 5. Looking at tables 1–4, we see that charging 5 points at the earliest iterations where $\boldsymbol{k}$ = 3 nets multiplier differences that are hugely larger than Equation 1 would yield. (e.g. table 3 has 914 and Table 1 shows 824, for a multiplier difference of 90 times.)

The money multiplier results of tables 1 through 4 also show why Equation 5 would motivate lower reserves in the context of Equation 5. (e.g. for $\boldsymbol{k}$ = 3, Table 4 shows 5,835 while Table 5 shows 5,232, for a multiplier difference of 673.) In a system wherein “too big to fail” results in public guaranteeing large scale failures, when the music stops the biggest jackpot goes to those who hit the wall highest.

In the context of such high multipliers, larger fee totals for transactions are achievable though that aspect is not examined herein. Additionally, interest on the new loans that were based on DIN capital is another source of income, and interest rate differentials (also not examined here) can potentially compensate for ongoing DIN costs in order to maintain a contract that is insuring the DIN which is the basis of a specific loan.

### 3.6 External Limits on Money Creation

Creditworthy borrowers are a necessary component of banking or loan service will fail. Some fraction of loans will go into default in any banking system. The question is how large that fraction is and whether the rate of default is supportable. In the GFC, limits were exceeded when too many loans defaulted for real estate purchases (The Financial Crisis Inquiry Commission 2011). It is clear from this that in the GFC, money multiplication hit external limits. That limits are present is also intuitively obvious. Such mechanisms cannot operate forever in the real world.

The modern economy may be primarily credit-driven, which correlates with the observation that fiat money creation lags credit creation, at least in boom times, contrary to the monetarist view (Keen 2009a). A related argument says that in times of contraction, fiat money introduced into the system will not get expanded by the banking multiplier because the system's contraction is not a result of lack of fiat money in the first place (Keen 2009b). Since banks are allowed to back-fill their needs for fiat money as needed and institutions are reluctant to cause a credit crunch by refusing these requests, loan activity drags fiat money along to keep pace with private money creation.

That a push relationship to money creation does not exist also makes intuitive sense because it is obvious that demand for loans from borrowers' drives loan activity and that banks should loan to those considered good credit risks. Thus, providing money to borrowers in some fashion improves borrower creditworthiness. (Whether the evaluation of credit risk by banks, both positive and negative, is correct is another question, and as I discuss above, there are pathological incentives present today.)

The monetarist view has treated the GFC as a signal for heroic extension of money to banks by the Federal Reserve in order to stimulate lending. In the monetarist view doing so provides the fuel to allow the banking system's multiplier to make credit available. This view is correct, but only to the extent that credit-worthy borrowers are available.

From both viewpoints, the existence of the Kraken multiplier for a segment of the economy is significant. In the monetarist view, it enables a far greater amount of credit to be created than would normally be thought possible under the currently understood money multiplier, with significant impacts on bailout requirements. In the credit demand driven view that eschews the push relationship on credit creation, it would be expected to raise the ratio of private debt to fiat money in unexpected ways.

Both views should see that this new multiplier applied to speculative (i.e. Ponzi, or price kiting) debt will fuel a greater capacity for "irrational exuberance". Both camps should realize that the existence of this new multiplier throws existing monitoring ideas into a cocked hat. Monitoring demand from banks if this new multiplier is operating cannot be a guide to credit extension activity as a rule of thumb when the Kraken is in play, and money ratios could change in ways hard to explain by normal mechanisms.

## 4 Concluding Comments

The Kraken equation can provide banks that use it with an advantage over other banks as long as the seller of default insurance notes (DINs) is able to cover liabilities or else have buyer liabilities guaranteed by the public to prevent the chaos a banking crash would cause. In theory, it can also be highly productive. In practice, using this method to create new lending capital when no commensurate creation of new utilitarian value can reasonably be expected is similar to a pyramid scheme in that it simply cannot go on forever. Although in theory brakes should be applied to availability of DINs when rating agencies downgrade the ratings of underlying securities, in practice they were not.

Complicating this are certain pathological incentives toward economically destructive banking practices. Profiting more by an unsound loan than from a properly made loan because of making a failure bet with a CDS is problematic for banking. It presents special concerns for Kraken banking when such synthetic capital creation is judged acceptable. Additionally, the current legal climate may also incentivize holders of CDS instruments to be uninterested in preventing bankruptcy of counterparties.

This mechanism, generally, presents issues of significant interest to regulators and in some cases may justify loosening or tightening reserve requirements for specific banks. It would be desirable at minimum to improve transparency, as well as to regulate disclosure of whether a CDS is an original insurer or part of a chain insurance, and provide information as to all parties in any chain. It may also be desirable to incentivize counterparties against bankruptcy of a financial entity by controlling the practice of first-position collections for certain derivatives like CDS's. At least some degree of due process is probably needed regarding taking immediate possession of cash collateral as opposed to property.

It would also be wise to take steps to end the practice of buying a CDS for a security which the buyer may know to be problematic. One possible method could be to require that CDS's expire when ownership of an underlying security is transferred, or perhaps require ownership transfer of the CDS to occur together with the underlying security. Non-owner CDS acquisition could be separated from the banking industry or require special registration and disclosure to sellers. It could be desirable also that speculative buyers of CDS instruments be put in intermediate position at bankruptcy and not fully covered by safe-harbor provisions. At minimum, where DIN based synthetic capital is created, it could be required that the DIN be bound to the security on which it is based so that it always transfers ownership together with the underlying security. A last measure could be to perhaps have a level of default above which the safe-harbor provisions would be nullified.

There is no inherent theoretical reason why Kraken banking should not be acceptable as long as the underlying securities are sound and everyone understands what is going on. It is conceivable that should an economy have valid needs for monetary expansion based on underlying productive fundamentals that the Kraken mechanism may be useful, perhaps even necessary, in order to provide an optimum amount of money within the economy. The advent of Kraken banking could prove to be positive by making availability of private money more dependent on unfettered human ability to judge commercial prospects for value creation in the absence of the fiscal limitations in the reserve banking system.

As part of an overall system oriented toward speculative transactions, Kraken banking is clearly dangerous. The issues raised guarantee that it could not be metastable under current regulation. Allowing its use by banks should be done with care toward the aspects of the economy it is fueling, as it has the ability to potentiate serious financial bubbles and there are overwhelming financial incentives to maximize short-term gain at public expense by using Kraken banking in a “too big to fail” financial system.

## Glossary of Terms:

**AIG –** American International Group corporation.
**BIS** – Bank for International Settlements.
**CDS** – Credit default swap.
**CRA** – Credit rating agency (Ex. Moody's)
**DIN** – Default insurance note. A type of insurance that pay the balance on a loan or investment should it fail. A generic term covering credit default swap (CDS) and other types of insurance on financial instruments. (The term DIN was coined as a generic for this publication.)
**GFC** – Global financial crisis.
**Kraken - A** mythical beast from Scandinavian folklore with many tentacles that rises without warning from the depths of the ocean to drag ships to their doom. It has appeared (incorrectly) in movies as a creature of Zeus, god of Mount Olympus. In this usage it is metaphorical.
**SEC** – Securities and Exchange Commission.
**tranche** – When investments in a portfolio are grouped together, rules can be designed such that a part of the portfolio is paid off by different rules than other parts. For instance, a first tier tranche can be paid off before all others, second tier after the first, etc.. Each of these parts is a tranche and carries a different level of risk relative to the pool of investments. Typically, the risk level of a tranche is reflected in the interest rate paid.

## Acknowledgements

This paper is dedicated to the late Kevin Walsh, Dean's Executive Professor of Management for the Leavey School of Business at Santa Clara University, venture capitalist with Ridge Partners and friend. His intelligence, wry humor and kindness will be missed. Thanks are also given to Jim Rutt, Steve Keen and John Quiggin for early comments.

Please note:

You are most sincerely encouraged to participate in the open assessment of this article. You can do so by either recommending the article or by posting your comments.

Please go to:

The Editor